\documentclass[prd,twocolumn,groupedaddress,showpacs,nofootinbib]{revtex4}
\usepackage{graphicx}
\usepackage{dcolumn}
\usepackage{amssymb}
\usepackage{mathrsfs}
\usepackage{amsmath}
\usepackage{epsfig}
\usepackage[dvips]{color}
\usepackage{hhline}

\begin{document}

\title{Electrophilic dark matter with dark photon: from DAMPE to direct detection}

\author{Pei-Hong Gu$^1$}
\email{peihong.gu@sjtu.edu.cn}

\author{Xiao-Gang He$^{2,3,4}$}
\email{hexg@sjtu.edu.cn}

\affiliation{$^1$School of Physics and Astronomy, Shanghai Jiao Tong
University, 800 Dongchuan Road, Shanghai 200240\\
$^2$T-D Lee Insitute, Shanghai Jiao Tong University, 800 Dongchuan Road, Shanghai 200240\\
$^3$Department of Physics, National Taiwan University, Taipei 106\\
$^4$National Center for Theoretical Sciences, Hsinchu 300}

\begin{abstract}

The electron-positron excess reported by the DAMPE collaboration recently may be explained by an electrophilic dark matter (DM).  A standard model singlet fermion may play the role of such a DM when it is stablized by some symmetries, such as a dark $U(1)_X^{}$ gauge symmetry, and dominantly annihilates into the electron-positron pairs through the exchange of a scalar mediator. The model, with appropriate Yukawa couplings, can well interpret the DAMPE excess. Naively one expects that in this type of models the DM-nucleon cross section should be small since there is no tree-level DM-quark interactions. We however find that at one-loop level, a testable DM-nucleon cross section can be induced for providing ways to test the electrophilic model. We also find that a $U(1)$ kinetic mixing can generate a sizable DM-nucleon cross section although the $U(1)_X^{}$ dark photon only has a negligible contribution to the DM annihilation. Depending on the signs of the mixing parameter, the dark photon can enhance/reduce the one-loop induced DM-nucleon cross section.

\end{abstract}


\maketitle

\section{Introduction}

Recently the DAMPE collaboration \cite{dampe2017} has reported a direct measurement of the high-energy cosmic-ray electrons and positrons in the energy range $25\,\textrm{GeV}$ to $4.6\,\textrm{TeV}$ with unprecedentedly high energy resolution and low background. While large part of the spectrum is well fitted by a smoothly broken power-law model, the spectrum seems to have a narrow bump above the background at around $1.4\,\textrm{TeV}$ \cite{dampe2017}. This excess indicates the existence of a new source of primary electrons and positrons.

It has been known that the annihilation of the dark matter (DM) particles can be a possible origin for the primary cosmic rays \cite{tw1990}. If the DM particles are expected to account for the DAMPE excess, they should mostly annihilate into the electron-positron pairs while the other leptonic and hadronic processes should be suppressed. It seems not easy to realize such an electrophilic DM annihilation through the mediation of certain gauge bosons which usually couple to both electron and other lepton flavors or not couple to electron \cite{bhy2009}. Alternatively, a Majorana \cite{knt2003,ma2006} or Dirac \cite{bglz2009,cms2009} DM fermion pair can annihilate into a lepton pair through the $t$-channel exchange of a singly charged mediator scalar. If the DM is a scalar, it can couple to the usual leptons with some singly charged mediator leptons \cite{cehl2014}. By choosing the related Yukawa couplings, such a leptophilic DM could explain the DAMPE excess. Naively, it would be difficult to directly detect such electrophilic DM particles since they have no tree-level couplings to quarks and therefore hadrons.

In this paper we shall construct an electrophilic DM model in which a DM fermion is stablized by a dark $U(1)_X$ gauge group and its Yukawa coupling to a mediator scalar and the standard model (SM) electron works for explaining the DAMPE data. The electrophilic nature of the DAMPE data is explained by the dominance of the Yukawa interaction. As long as the $U(1)_X^{}$ gauge coupling is small enough and/or the dark photon is heavy enough, the DM annihilation involving the dark photon will only give a negligible contribution to the total DM annihilation and hence will not soften the narrow bump observed in the DAMPE spectrum produced by an electrophilic interaction. We then evaluate the DM-nucleon scattering cross section. First of all, the Yukawa coupling among the DM, the mediator and the electron can always lead to a DM-nucleon scattering at one-loop level. On the other hand, since the dark photon couples to the SM fermions through a $U(1)$ kinetic mixing, it can generate a DM-nucleon scattering at tree level. The tree-level DM-nucleon scattering cross section can arrive at a testable level and can enhance/reduce the one-loop DM-nucleon cross section, depending on the signs of the mixing parameter. Therefore, the DM direct detection experiments can also provide useful information about the models for the DMAPE excess.

This paper is organized as follows. In Sec. II, we introduce the generally electrophilic models. In Sec. III, we focus on a simple model to demonstrate the electrophilic DM annihilations and scatterings. Finally, we make a conclusion in Sec. IV.

\section{An electrophilic DM model with dark photon}

The models of the type we have in mind can be classified into two types: the type-A models with mediator scalar and DM fermion, and the type-B models with mediator fermion and DM scalar. If the mediator and DM fields couple to the SM lepton doublets, they may be related to the neutrino mass generation. In this case, it may be difficult to realize an electrophilic DM because of the constraint from the neutrino flavor mixing. Therefore we shall only consider the couplings of the mediator and DM fields to the SM right-handed leptons. The type-A models then can contain the mediator scalar and the DM fermions as below, 
\begin{eqnarray}
\textrm{Mediator~scalar}:&&\delta(1,1,-1)(-1)\,,\nonumber\\
[2mm]
\textrm{DM~fermion}:&&\chi_{L,R}^{}(1,1,0)(+1)\,.
\end{eqnarray}
As for the type-B models, their mediator fermions and DM scalar should be
\begin{eqnarray}
\textrm{Mediator~fermion}:&&E_{L,R}^{}(1,1,-1)(-1)\,,\nonumber\\
[2mm]
\textrm{DM~scalar}:&&\chi(1,1,-1)(+1)\,.
\end{eqnarray}
Here and thereafter the first and second brackets describe the transformation under the SM $SU(3)_c^{}\times SU(2)_L^{} \times U(1)^{}_{Y}$ gauge symmetry and the dark $U(1)_X^{}$ gauge symmetry. 

For demonstration, we shall focus on the following type-A model,  
\begin{eqnarray}
\label{lar}
\mathcal{L}&=&\mathcal{L}_{\textrm{SM}}^{} -\frac{\epsilon}{2}B_{\mu\nu}^{}X^{\mu\nu}-\frac{1}{4}X_{\mu\nu}^{}X^{\mu\nu}\nonumber\\
[2mm]
&&+i\bar{\chi}_{L,R}^{}\gamma^\mu_{} D_\mu^{} \chi_{L,R}^{} -m_\chi^{}\left(\bar{\chi}_L^{}\chi_R^{} +\textrm{H.c.}\right)\nonumber\\
[1mm]
&&-\sum_{\alpha=e,\mu,\tau}^{}\left(f_{\alpha}^{}\delta \bar{e}_{R\alpha}^{} \chi_L^{}+\textrm{H.c.}\right )+(D_\mu^{}\xi)^\dagger_{}D^\mu_{}\xi\nonumber\\
[1mm]
&& - \left(\mu_\xi^2 +\lambda_{\xi\phi}^{}\phi^\dagger_{}\phi\right)\xi^\dagger_{}\xi -\lambda_\xi^{}  \left(\xi^\dagger_{}\xi\right)^2_{}+(D_\mu^{}\delta)^\dagger_{}D^\mu_{}\delta\nonumber\\
[2mm]
&&- \left(\mu_\delta^2 +\lambda_{\delta\xi}^{}\xi^\dagger_{}\xi+\lambda_{\delta\phi}^{}\phi^\dagger_{}\phi\right)\delta^\dagger_{}\delta - \lambda_\delta \left(\delta^\dagger_{}\delta \right)^2_{}\,,
\end{eqnarray}
where $\mathcal{L}_{\textrm{SM}}^{}$ denotes the SM Lagrangian, $B_{\mu}^{}$ and $X_\mu^{}$ are the $U(1)_Y^{}$ and $U(1)_X^{}$ gauge fields, $\xi(1,1,0)(-1)$ is a Higgs scalar for spontaneously breaking the $U(1)_X^{}$ gauge symmetry, $\phi(1,2,+\frac{1}{2})(0)$ and $e_R^{}(1,1,-1)(0)$ are the SM Higgs doublet and lepton singlets. 
The covariant derivatives $D_\mu^{}$ are as below,
\begin{eqnarray}
&&D_\mu^{}\xi=\left(\partial_\mu^{} +i g_X^{}  X^{}_{\mu} \right) \xi\,,\nonumber\\
[2mm]
&&D_\mu^{}\delta=\left(\partial_\mu^{} + i g_X^{} X^{}_\mu  + i g' B^{}_{\mu}\right) \delta \,,\nonumber\\
[2mm]
&&D_\mu^{}\chi_{L,R}^{}=\left(\partial_\mu^{} - i g_X^{} X^{}_{\mu} \right) \chi_{L,R}^{}\,, 
\end{eqnarray}
with $g_X^{}$ and $g'$ being the $U(1)_X^{}$ and $U(1)_Y^{}$ gauge couplings.

Remarkably that the dark $U(1)_X^{}$ gauge symmetry in the model of Eq. (\ref{lar}) has forbidden other possible Yukawa couplings of the dark fermion $\chi$ to the SM fields. As a consequence, the dark fermion $\chi$ is kept stable to serve as a DM particle if it has a right relic density \cite{patrignani2016}.

After the dark Higgs scalar $\xi$ develops its vacuum expectation value (VEV),
\begin{eqnarray}
\xi=\frac{1}{\sqrt{2}}\left(v_\xi^{}+h_\xi^{}\right),
\end{eqnarray}
to spontaneously break the dark $U(1)_X^{}$ gauge symmetry, the dark photon $X_\mu^{}$ can acquire a mass,
\begin{eqnarray}
m_X^2 = g_X^2 v_\xi^2\,.
\end{eqnarray}
As for the SM Higgs doublet $\phi$, it is responsible for the electroweak symmetry breaking as usual, i.e.
\begin{eqnarray}
\phi=\left[\begin{array}{c}0\\
[2mm]
\frac{1}{\sqrt{2}}\left(v_\phi^{}+h_\phi^{}\right)
\end{array}\right].
\end{eqnarray} 
Then the mass of the mediator scalar $\delta$ is easy to read,
\begin{eqnarray}
M_\delta^2 = \mu_\delta^2 + \frac{1}{2}\lambda_{\delta\xi}^{}v_\xi^2 + \frac{1}{2}\lambda_{\delta\phi}^{}v_\phi^2\,.
\end{eqnarray}
Due to the $U(1)$ kinetic mixing, the dark photon $X_\mu^{}$ can couple to the SM fermion pairs, i.e.
\begin{eqnarray}
\mathcal{L}\supset \epsilon X_\mu^{} \left(-\frac{1}{3}\bar{d}\gamma^\mu_{}d +\frac{2}{3}\bar{u}\gamma^\mu_{}u -\bar{e}\gamma^\mu_{}e \right)~~\textrm{for}~~\epsilon \ll 1\,. \label{dphoton-fermion}
\end{eqnarray}  
The dark photon $X_\mu^{}$ is no longer a mass eigenstate because of the $U(1)$ kinetic mixing. However, for a small $U(1)$ kinetic mixing, i.e. $\epsilon \ll 1$, the dark photon $X_\mu^{}$ can well approximate to a mass eigenstate.

\section{DM annihilation and scattering}

Exchanges of both the mediator scalar $\delta$ and the dark photon $X_\mu^{}$ can annihilate DM into SM particles. Since the dark photon $X_\mu^{}$ couples to all SM fermions with similar strength, it would produce a final state to soften the bump in the electron-positron spectrum observed by the DAMPE satellite. Therefore, the Yukawa couplings in Eq. (\ref{lar}) are expected to dominate the DM annihilation with $f_e^{}$ being much larger than $f_{\mu,\tau}$. Through the $t$-channel exchange of the mediator scalar $\delta$, the dark fermion $\chi$ can annihilate into the lepton pairs. The cross section is given by 
\begin{eqnarray}
\langle\sigma\left(\chi\bar{\chi}\rightarrow e_{R\alpha}^{}\bar{e}_{R\beta}^{}\right)v_{\textrm{rel}}^{}\rangle  = \frac{\left|f_{\alpha}^{}f_{\beta}^\ast\right|^2_{}}{16\pi}\frac{m_\chi^2}{\left(m_\chi^2+M_\delta^2\right)^2_{}}\,.
\end{eqnarray}
In addition, the dark fermion $\chi$ can annihilate into two dark photons $X_\mu^{}$ or into one dark photon $X_\mu^{}$ and one dark Higgs boson $h_\xi^{}$. We calculate the cross section by 
\begin{eqnarray}
\!\!\!\!\langle\sigma\left(\chi\bar{\chi}\rightarrow XX \right)v_{\textrm{rel}}^{}\rangle &=& \frac{g_X^4}{16\pi}\frac{1}{m_\chi^2} ~~\textrm{for}~~m_\chi^{2}\gg m_X^{2}\,,\nonumber\\
[2mm]
\!\!\!\!\langle\sigma\left(\chi\bar{\chi}\rightarrow Xh_\xi^{} \right)v_{\textrm{rel}}^{}\rangle &=& \frac{3g_X^4}{256\pi}\frac{m_X^2}{m_\chi^4} ~~\textrm{for}~~m_\chi^{2}\gg m_{X,h_\xi^{}}^{2}\,.\nonumber\\
\!\!\!\!&&
\end{eqnarray} 
Furthermore, the dark photon $X_\mu^{}$ can mediate the annihilation of the dark fermion $\chi$ into the SM fermion pairs as shown in Eq. (\ref{dphoton-fermion}) because of the $U(1)$ kinetic mixing. Clearly, this annihilation channel should be negligible when the dark photon is heavy and/or the kinetic mixing parameter $\epsilon$ is small.

In order to explain the positron anomaly reported by the DAMPE, the dark fermion should mostly annihilate into the electron-positron pairs. For this purpose, we can take the Yukawa couplings $f_{e,\mu,\tau}^{}$ to be 
\begin{eqnarray}
f_{e}^{}\gg f_{\mu,\tau}^{}\,.
\end{eqnarray}
This choice may be understood by additional symmetries such as the Froggatt-Nielsen mechanism \cite{fn1979}.  We need also to take the $U(1)_X^{}$ gauge coupling $g_X^{}$ to be smaller than the Yukawa coupling $f_e^{}$ if the dark photon is not heavy. Otherwise, the dark fermion pairs will mostly annihilate into two dark photons, which subsequently decay into the SM fermion pairs. In this case, the dark fermion annihilation will give a wider bump in the electron-positron spectrum even if the dark photon is light enough and hence only decays into the electron-positron pairs.  

We now consider further test of the model from the DM direct detections. Naively one expects that in the type of models discussed above the DM-nucleon cross section should be small since there is no tree-level DM-quark interactions, we however find that this is not completely true due to two reasons. On one hand, the Yukawa couplings in Eq. (\ref{lar}) will lead to a coupling of the dark fermion to a virtual photon and hence to the quark current. On the other hand, the dark fermion can scatter off the protons in nuclei at tree level through the $t$-channel exchange of the dark photon in the presence of the $U(1)$ kinetic mixing. These contributions to the DM-nucleon cross section need to be studied more carefully. By taking into account both of the mediation of the real dark photon and the virtual ordinary photon, the effective operators at quark level for the scattering should be \cite{bth2011}
\begin{eqnarray}
\mathcal{L}&\supset& a_\gamma^{}Q_q^{}\bar{q}\gamma^\mu_{}q \bar{\chi}\gamma_\mu^{} P_L^{}\chi + a_X^{} Q_q^{}\bar{q}\gamma^\mu_{}q \bar{\chi}\gamma_\mu^{} \chi~~\textrm{with}\nonumber\\
[2mm]
&&a_{\gamma}^{}= \frac{e^2_{}\left|f_\alpha^{}\right|^2_{}}{16\pi^2_{} M_\delta^2} \left[\frac{1}{2} +\frac{1}{3}\ln\left(\frac{m_{\alpha}^2}{M_\delta^2}\right)\right]  \,,~~a_X^{}= \frac{ \epsilon e g_X^{} }{m_X^2}  \,,\nonumber\\
&&
\end{eqnarray}
with $Q_q^{}$ being the electric charge of the quark $q$. In the above $a_\gamma$ is induced by the exchanges of the mediator scalar $\delta$ and the electron at one-loop level while $a_X$ is generated by the exchange of the dark photon $X_\mu^{}$ at tree level.

The DM-nucleon scattering cross section is given by
\begin{eqnarray}
\sigma_{\chi p\rightarrow \chi p}^{}= {\left(a_X^{}+\frac{1}{2}a_\gamma^{}\right)^2_{}  \over 2\pi} 
{m^2_\chi m^2_p \over \left(m_\chi ^{}+m_p^{} \right)^2_{}}\,.
\end{eqnarray}
From the above, one sees that the total DM-nucleon cross section is a sum of the one-loop and tree amplitudes squared. If their sizes are comparable, depending on the signs of the kinetic mixing parameter $\epsilon$, the total cross section can be either enhanced or reduced. The DM direct detection limit can be used to constrain the model parameters.

We find that the parameter space of the model can allow the possibility that the DM annihilation due to the exchange of the mediator scalar $\delta$ produces the correct relic density, and to explain the DAMPE electron-positron excess. For example, by setting the following parameters, 
\begin{eqnarray}
\label{par}
&&m_\chi^{}=1.4\,\textrm{TeV}\,,~~M_\delta^{}=7\,\textrm{TeV}\,,~~m_X^{}\ll \,m_\chi^{}\,,\nonumber\\
[2mm]
&&f_{e}^{}=\sqrt{4\pi}\gg f_{\mu,\tau}^{}\,, ~~g_X^{}=0.1\,,
\end{eqnarray}
we obtain the annihilations from the mediator scalar $\delta$ exchange and the dark photon$X_\mu^{}$ exchange to be
\begin{eqnarray}
\langle\sigma\left(\chi\bar{\chi}\rightarrow e^{-}_{}e^{+}_{}\right)v_{\textrm{rel}}^{}\rangle &\simeq& 1\,\textrm{pb}\,,\nonumber\\
[2mm]
\langle\sigma\left(\chi\bar{\chi}\rightarrow Xh_\xi^{} \right)v_{\textrm{rel}}^{}\rangle&\ll&\langle\sigma\left(\chi\bar{\chi}\rightarrow XX \right)v_{\textrm{rel}}^{}\rangle \nonumber\\
[2mm]
&\simeq& 4\times 10^{-4}_{}\,\textrm{pb}\,.
\end{eqnarray}
The observed DM relic density \cite{patrignani2016} can be generated by the $\delta$ exchange, other than the $X_\mu^{}$ exchange. The contributions from the DM annihilation to the SM fermion pairs are proportional to $\epsilon^2$ and is tiny too.

Although the dark photon $X_\mu^{}$ has a negligible contribution to the DM annihilation, it can mediate a testable DM-nucleon scattering cross section \cite{lcj2017}. For the parameter choice (\ref{par}), we have
\begin{eqnarray}
a_X^{}&=& 3\times 10^{-3}_{}\,\textrm{TeV}^{-2}_{}\left(\frac{\epsilon}{10^{-3}}\right)\left(\frac{100\,\textrm{GeV}}{m_X^{}}\right)^2_{},\nonumber\\
[2mm]
a_\gamma^{}&=&-1.5\times 10^{-3}_{}\,\textrm{TeV}^{-2}_{},\nonumber\\
[2mm]
\sigma_{\chi p\rightarrow \chi p}^{}&=&1.4\times 10^{-46}_{}\,\textrm{cm}^2_{}\,.
\end{eqnarray}
The DM-nucleon cross section is  below the direct DM detection limit~\cite{dark} if the kinetic mixing parameter $\epsilon$ is of the order of $10^{-3}$ and the dark photon mass $m_X^{}$ is of the order of $100\,\textrm{GeV}$. Improved DM direct search can probe more parameter space. There are strong constraint on the dark photon parameter $\epsilon$ at low masses. However, with a dark photon mass of the order of $100\,\textrm{GeV}$, the constraint on $\epsilon$ has not ruled out its value to be of order $10^{-3}$ which may be tested at future colliders \cite{he-dark}.

We also check the contribution from the dark photon $X_\mu^{}$ to the muon magnetic dipole moment \cite{bdhk2001}, 
\begin{eqnarray}
\Delta a_\mu^{}&=&\frac{\epsilon^2 g_X^2}{12\pi^2_{}}\frac{m_\mu^2}{m_X^2}\nonumber\\
&\simeq&10^{-16}_{} \left(\frac{\epsilon}{10^{-3}}\right)^2_{}\left(\frac{g_X^{}}{0.1}\right)^2_{} \left(\frac{100\,\textrm{GeV}}{m_X^{}}\right)^2_{}\,.
\end{eqnarray}
The contribution is too small to play any significant role in explaining the g-2 anomaly of muon.

\section{Conclusion}

In this paper we have constructed a simple model with DM electrophilic interaction which can explain the electron-positron excess reported by DAMPE and have testable effects by DM direct detection experiments through kinetic mixing with a dark photon. A SM singlet fermion stablized by a dark $U(1)_X^{}$ gauge symmetry plays the role of the DM with a mass of $1.4\,\textrm{TeV}$. The dominant DM annihilation is from the Yukawa couplings of the DM to the leptons with a scalar mediator. We find that at one-loop level, a testable DM-nucleon cross section can be induced providing ways to test the electrophilic model. We also find that a $U(1)$ kinetic mixing can generate a sizable DM-nucleon cross section although the $U(1)_X^{}$ dark photon only has a negligible contribution to the DM annihilation. Depending on the signs of the mixing parameter, the dark photon can enhance/reduce the one-loop induced DM-nucleon cross section. The contribution to $g-2$ of muon is too small to play any role in explaining the $g-2$ anomaly of muon. 

Finally we would like to make a comment on the possibility that a slightly soft peak of electron-positron spectrum may also fit the DAMPE excess well when the DM annihilates into the electron-positron pairs and the muon-antimuon pairs with $1:1$ ratio. In this case the $U(1)_{L_e-L_\mu}^{}$ leptophilic DM model discussed in Ref. \cite{bhy2009} may provide an excellent scenario to explain the DAMPE excess, with the $U(1)_{L_e-L_\mu}^{}$ gauge boson $Z'$ not too far from the DM pair resonant point.

\textbf{Acknowledgement}: 
This work was supported in part by the Key Laboratory for Particle Physics, Astrophysics and Cosmology, Ministry of Education, and the Shanghai Key Laboratory for Particle Physics and Cosmology (Grant No. 15DZ2272100). P.H.G. was supported in part by the NSFC (Grant No. 11675100) and also in part by the Recruitment Program for Young Professionals (Grant No. 15Z127060004). X.G.H. was supported in part by the NSFC (Grant Nos. 11575111 and 11735010) and also in part by the MOST (Grant No. MOST104-2112-M-002-015-MY3).

\end{document}